# Analysing the factors affecting electric vehicle adoption using the extended theory of planned behaviour framework

Authors: Pranshu Raghuvanshi[1] and Anjula Gurtoo[1]

## Abstract

This study uses the Theory of Planned Behaviour (TPB) framework and expands it by including Optimism, Innovativeness and Range Anxiety constructs. In this study, conducted in Lucknow - the capital of India's most populous province (Uttar Pradesh)– a multi-stage random sampling design was employed to select 432 respondents from different city areas. The survey instruments were adapted from similar studies and suitably modified to suit the context. Using exploratory factor analysis, 18 measurement items converged into six factors: attitude, subjective norms, perceived behavioural control, optimism, innovativeness and range anxiety. We confirmed the reliability and validity of the constructs using Cronbach's alpha, composite reliability, average variance extracted and discriminant validity analysis. We explored the relationship between them using structural equation modelling. All factors but Optimism were found to be significantly associated with adoption intention. We further employed mediation analysis to examine the mediation pathways. The TPB components mediated the effect of innovativeness but not range anxiety. The study's insights can help policymakers and marketers design targeted interventions that address consumer concerns, reshape consumer perceptions, and foster greater EV adoption. The interventions can target increasing the mediating variables or decreasing range anxiety to facilitate a smoother transition to sustainable transportation.

**Highlights**

1. The study focuses on the adoption of EVs in an Indian tier II city. Most EV studies in India concentrate on metros; the **implications** of the results are important in the context of tier-II cities, where rising income and rapid car ownership growth present significant EV adoption potential.

---

[1] Indian Institute of Science, Bangalore, India

2. The TPB components alone may not be sufficient to explain the EV adoption behaviour. The framework must be expanded to include context-specific psychological and technological factors.
3. Innovativeness and Range Anxiety (variables outside the TPB) affect the EV adoption intention. However, innovativeness influences EV adoption intention entirely through TPB components (attitude, subjective norms and perceived behavioural control). While range anxiety operates outside TPB and directly affects adoption intention, this **implies** that interventions should not only strengthen TPB components but also address range-related concerns independently.
4. Consumer optimism does not directly affect the EV adoption intention as postulated by the theories on technology adoption. However, it does shape the TPB's attitude and perceived behavioural control components. This **implies** that initiatives aimed at boosting optimism alone may not increase adoption intention unless paired with strategies that convert optimism into favourable attitudes and perceived control, such as awareness campaigns and hands-on EV experience.

**Introduction**

The transition towards clean technologies has made Electric vehicles (EVs) a key instrument in achieving sustainable mobility. Twenty-three per cent of the $CO_2$ emissions globally come from the transport sector, and three-quarters of it can be attributed to road traffic (IEA, 2023). In India, the transport sector accounts for sixteen per cent of the total $CO_2$ emissions, 90 per cent of which is attributed to road transport (IEA, 2023). The IEA also reports that the number of four-wheeled vehicles is increasing faster in developing countries. The increase in disposable income in India post-liberalisation and access to credit created a large market for the private four-wheeler vehicles (Ishaq & Mohideen, 2018). In the last decade, private motorised vehicles more than doubled, reaching 300 million in 2020, compared to 141 million in 2010 (MoRTH, 2020). This figure was projected to be reached in the year 2040 (Bhat, 2024). India became the fourth biggest market for automobiles in 2022, surpassing Germany (Wadhwa, 2022). Automobile sales are expected to further rise in the coming years (Bansal, Dua, Krueger, & Graham, 2021).

In such a scenario, accelerating EV adoption through subsidies and incentives, awareness campaigns and infrastructure development, like setting up charging stations, is a viable solution to address mounting concerns over climate change. However, despite these efforts, the adoption efforts are faced with challenges. The penetration of EVs in the private car segment is sluggish. Only 7.46 per cent of the new vehicles sold in 2024 were EVs (Kumar, 2025). Comparing this figure with the 30 per cent target set for 2030 under the FAME (Faster Adoption of Electric Vehicles) India scheme seems to be an underachievement. The consumers continue to face barriers to the adoption of EVs.

To address these barriers and design effective policies, it is necessary to understand the factors that directly impact EV adoption and mediate the relationship between other variables and adoption intention. As per the Theory of Planned Behaviour (TPB), Behavioural intention to adopt a product, idea, or technology depends on three factors: attitude, subjective norms, and perceived behavioural control (Azjen, 1991). The TPB is a widely used framework in behavioural research to study individuals' decision-making across domains like health, sustainability and technology adoption (Godin & Kok, 1996; Klöckner, 2013). While TPB is an excellent basic framework with universal applicability, context-specific variables significantly shape consumer behaviour. Recent research on technology adoption has extended the theory of planned behaviour to include context-specific variables like technological innovativeness, environmental concerns, moral values of the individual, etc. EV adoption faces unique technological and psychological barriers like range anxiety - fear of running out of charge before reaching the destination or charging station. Despite positive attitudes and norms, these barriers can deter EV adoption.

This study also incorporates additional constructs, reflecting unique concerns surrounding EV adoption, and extends the traditional TPB framework. This study uses empirical data to analyse the relative influence of core TPB components and extended constructs/factors in determining adoption intention. The study also checks for the mediating role of the core TPB components. It includes only those extended constructs in the final model whose effect on adoption intention is not entirely mediated by the core TPB components. The results offer a more nuanced understanding of consumer behaviour concerning EV adoption in tier 2 cities and provide practical insights to manufacturers, marketers, and policymakers for enhancing EV penetration. This study contributes to the literature on electric vehicle adoption in the following ways.

1. Firstly, we use the theory of planned behaviour (TPB) framework to study the intention to adopt EVs. However, we do not limit the study to the framework; instead, we include some psychological variables outside the theory and study their impact on the intention to adopt EVs. Therefore, we propose an extended TPB for the adoption of EVs where variables like innovativeness, range anxiety, and optimism impact either the components of the TPB and thereby indirectly or directly impact the intention to adopt EVs. We also explore the dependence among these variables using Structural Equation Modelling (SEM). Hence, we find specific psychological variables, including but not limited to the variables in the TPB, that impact EV adoption intention.
2. Secondly, unlike existing studies on EV adoption, which focus on tier-I cities like Bengaluru, Delhi, Mumbai, etc., this study was done in a tier–II city in India (Lucknow). There is immense scope for EV penetration in provincial capitals and other tier-II cities. As per the classification of cities by the Indian Government, the number of tier-II cities in India stands at 97 against only eight tier-I cities. With significant earnings of people in these cities, private car ownership is increasing rapidly. Between 2011 and 22, private car ownership in Lucknow increased 2.5 times (TOI, 2023). Therefore, to come up with more effective policies to reduce emissions, efforts to be made to increase EV adoption in tier-II cities. Hence, we systematically study the factors affecting EV adoption and suggest policies to improve it.

We contribute to the understanding of EV adoption by addressing these gaps in the literature. In the next section, we present a review of literature on the extended theory of planned behaviour (TPB) and EV adoption. This is followed by the description of the methodology in Section 3. Section 4 reports the results of the study, followed by a discussion in Section 5 and a conclusion in Section 6.

## 2. Literature Review
**Theory of Planned Behaviour**

Icek Ajzen, in the 1980s, came up with the "Theory of Planned Behaviour". The theory extends his previous work, the "Theory of Reasoned Action" (Fishbein & Ajzen, 1975). It postulates that human behaviour results from three factors: attitude, perceived behavioural control and subjective norms (Azjen, 1991). These factors contribute to the person's intention to engage in a particular behaviour. Intention is an individual's motivational readiness to perform a behaviour (Fishbein & Ajzen, 2011). The performance of actual behaviour is affected by the intention, which is determined by attitude, perceived behavioural control and subjective norms. Attitude pertains to the degree of an individual's favourable or unfavourable evaluation of a particular behaviour (Beck & Ajzen, 1991). A favourable assessment may result in the performance of the behaviour, while an unfavourable evaluation may result in non-performance. In this study, attitude pertains to an individual's beliefs, feelings, and perceptions about electric vehicles (Ajzen, 2020), which personal beliefs, experience, and increased knowledge can influence. Perceived Behavioural Control is the individual's assessment of their ability and capacity to perform the behaviour, given existing opportunities, challenges and constraints (Agyei, Asamoah, Nanor, & Boansi, 2025). It is the belief that controls behaviour like purchasing and using an electric vehicle (Ajzen, 1991). Perceived behavioural control is a novel addition and was not part of the original "Theory of Reasoned Action". Intention to perform a behaviour is also influenced by social pressure. Individuals conform to group norms and are part of various groups. The perceived influence of social groups affects their intention to perform the behaviour (Beck & Ajzen, 1991). This is termed as Subjective Norms in the theory of planned behaviour. Some behaviours are considered desirable by the group, and others are not. Individuals may be inclined to perform desirable behaviours and avoid performing undesirable ones.

The effect of the three core components of the TPB varies and depends on the product and context. In some studies, for example, the effect of attitude may dominate the other two and vice versa. A constructive attitude increases the likelihood of adoption of a product (Gupta, 2025). Therefore, attitude determines a person's inclination to buy an EV (Mohamed, Higgins, Ferguson, & Kanaroglou, 2016). The impact of attitude on the adoption intention of a sustainable product may vary in intensity. Some studies on green and sustainable adoption report a weak influence of attitude on adoption intention (Gupta A. K., 2021; Srivastava & Gupta, 2023). While some others found a strong impact of attitude on the adoption of EVs and other sustainable products (Gidaković, et al., 2024; Shakeel, 2022; D'Souza, Brouwer, & Singaraju, 2022). Similarly, the effect of subjective norms and perceived behavioural control has been found to vary in different contexts. Subjective norms were found to be more effective in studies on organic food consumption (Sadiq, Rajeswari, Ansari, & Kirmani, 2021) and solar rooftop PV adoption (Abreu, Wingartz, & Hardy, 2019). While the effect of attitude was weak, the effect of perceived behavioural control was insignificant. In the studies of Mohamed et al., Srivastava and Gupta, the effect of perceived behavioural control was the most effective component of the TPB, and the effects of the other two were weak. Since the effects of the three components are product and context-specific, it is essential to check their impact on EV adoption in a tier-II Indian city.

The components of the TPB are not independent of each other. They influence each other apart from affecting adoption intention. Individuals who consider EVs as socially desirable develop a positive attitude towards them (Singh, Singh, & Vaibhav, 2020). They will also have a higher perceived control over EVs when they see people in their social circle driving EVs. Literature suggests that if close people support a behaviour, then perceived behavioural control is higher (Putry & Harsono, 2021). Similarly, higher perceived behavioural control creates a positive attitude towards the behaviour (Ogiemwonyi & Harun, 2021). Therefore, in many studies, the three core components of the TPB are not found to be orthogonal to each other. They may or may not be significantly correlated with each other, as well as affect the EV adoption intention.

On the other hand, the three core components are not the only factors affecting adoption intention. Based on a study conducted in Canada, which included 3505 respondents, it was found that the willingness to purchase EVs was dependent on the person's attitude and perceived behavioural control, social norm and personal moral norm (Mohamed, Higgins, Ferguson, & Kanaroglou, 2016). The study extends TPB by including the personal moral norm in the framework. Similarly, Kaplan and others extend the framework by including the construct, 'perceived familiarity' with the EVs as a significant factor influencing their adoption (Kaplan, Gruber, Reinthaler, & Klauenberg, 2016). Vital factors like emotional responses to a product, which is are not fully captured by the core components of the TPB (Duran, Alzate, Lopez, & Sabucedo, 2011). Duran et al. extended the TPB to predict behaviour using emotional factors. Car driving emotions significantly drive the EV usage intentions (Moons & Pelsmacker, 2015).

Apart from the three core components of the theory of planned behaviour, we also considered other important psychological variables. The first two are innovativeness and optimism, which, according to the literature, drive behaviour and are instrumental in technology adoption (Salari, 2022). Innovativeness is the tendency to adopt new products and innovations (Morton, Anable, & Nelson, 2016). Innovative consumers tend to be technology pioneers (Goldsmith & Hofacker, 1991). Morton and others found that consumers wanting to own new technology preferred electric vehicles (Morton, Anable, & Nelson, 2016). EV, a new technology, motivates consumers (Heffner, Kurani, & Turrentine, 2007). Goldsmith and Hofacker also concluded that innovativeness is a personality trait found in everyone to some extent (Goldsmith & Hofacker, 1991). Hence, we hypothesise that Innovativeness affects the components of the "Theory of Planned Behaviour" (TPB).

Optimism is a positive view of technology (Parasuraman & Colby, 2015). It is a belief in technology that technology leads to flexibility, efficiency, and control in people's lives. The literature on technology adoption suggests a higher probability of adoption when optimism towards the technology is high (Parasuraman, 2000). The chances of adoption increase when consumers believe they can control technology, leading to flexibility in their lives. (Liljander, Gillberg, Gummerus, & Riel, 2006). Hence, optimism may be related to perceived behavioural control. It is also expected that optimistic people will possess a favourable attitude towards electric vehicles (Salari, 2022) and, therefore, will be more likely to adopt electric vehicles (Jensen, Cherchi, & Lindhard, 2013; Egbue & Long, 2012; Moons & De Pelsmacker, 2012). On the other hand, the role of optimism in EV purchase intention is not well established. Liljander and others found that optimism is unrelated to EV purchase intention. Therefore, we have considered optimism as a variable to evaluate its impact on attitude, perceived behavioural control and the electric vehicle adoption intention.

EV range is another essential attribute identified as a critical barrier to adopting electric vehicles (She, Sun, Ma, & Xie, 2017; Patyal, Kumar, & Kushwah, 2021). Limited range triggers anxiety among consumers, a fear that battery depletion during trips may not enable them to meet their travel needs (Fan, Chen, & Zhao, 2022; Wang & Deng, 2019; Zhang, et al., 2018). With technological advancements, the performance of electric vehicles has improved significantly (Junquera, Moreno, & Álvarez, 2016). Improved battery capacity (Li, et al., 2018), less charging time (Byun, Shin, & Lee, 2018), and improvements in charging infrastructure (Wang, Li, & Zhao, 2017) have addressed the issue of range anxiety to a considerable extent. However, due to informational asymmetry and inertial thinking, consumers have a strong prejudice against the EV (Raimi & Leary, 2014). Moreover, it is also seen in some empirical studies that the factors contributing to range anxiety are psychological (Wang, Chen, & He, 2025). Hence, it is not the actual range of the vehicle but the perception of the limited range that impacts the intention to adopt EVs. Consumers with a negative perception of the EV range may be less likely to adopt electric vehicles.

While researchers have analysed EV adoption using the TPB, the variables considered are different in studies due to their objectives and contextual differences. There also remains a lack of clarity in the literature regarding the effect of the core variables on the EV adoption intention. As already discussed, whereas some studies argue a stronger role of these variables on sustainable behaviour, others find a weak or even insignificant effect of some variables (Abreu, Wingartz, & Hardy, 2019; Sadiq, Rajeswari, Ansari, & Kirmani, 2021). It is also evident that some important concerns exist that operate outside the theory of planned behaviour. Therefore, to address these concerns appropriately, the theory should be suitably adapted after examining the mediating effect of the core components of the TPB. The correlation between core components of the TPB is also not explored enough in the adoption behaviour studies, especially those related to EV adoption. This lack of coherent understanding in the literature exists as a potential gap which needs to be filled to advance the understanding of EV adoption. Therefore, further investigation is necessary to conceptualise and empirically test an extended TPB model that incorporates these context-specific determinants in a structured and theoretically grounded manner. Hence, we propose the following hypothesis.

*$H_1$: Core components of TPB affect EV adoption intention.*
*$H_2$: Attitude, subjective norms, and perceived behavioural control are significantly correlated in the context of electric vehicle adoption.*
*$H_3$: Consumer optimism directly affects the electric vehicle adoption intention.*
*$H_4$: Innovativeness directly affects the electric vehicle adoption intention.*
*$H_5$: Range Anxiety directly affects the electric vehicle adoption intention.*
*$H_6$: TPB components mediate the relationship between external constructs and adoption intention.*

## 3. Methodology

This section briefly describes our data and sample, followed by the statistical techniques used for analysing the data.

### 3.1 Data and Sample

The data comes from a survey conducted in 2024 in a tier 2 city (Lucknow) in India. We chose to do it in a Tier 2 city after consulting experts in the market. Given the high-income level in a tier 1 city, consumers often own two vehicles, where EVs are a choice for the second vehicle. We wanted to study the preference of EVs as a first vehicle, which is possible in a tier 2 city. Also, Lucknow is an emerging market for EVs. Between 2011 and 22, private car ownership has increased 2.5 times (TOI, 2023). Given the provincial government's launch of the City Electric Mobility Plan (CEMP) to support the transition to electric mobility, Lucknow emerged as an appropriate location for this study (TOI, 2023).

The first part of the survey includes measurement items that measure respondents' perception of EV. The second part comprises socio-demographic and vehicle ownership, preference information, and intention to adopt EVs. Relevant literature informed the development of the measurement items. The measurement items were based on similar studies done in the past and adapted as per the requirements of the research and demography. The scales are not adopted as they are. We took measurements on a "five-point Likert scale". We collected 432 responses targeting individuals who were potential buyers of electric vehicles. After excluding incomplete responses and respondents below 18, we retained 407 responses for our analysis.

**Table 1** presents the demographic information of the respondents. The demographic data indicates that the respondents are potential buyers of 4-wheeled vehicles. Less than 10 per cent of the respondents reported an annual income of less than 500000 INR. More than 90 per cent have a family size of 3-4 or more. More than 85 per cent own a car. The sample's education level and age distribution are also appropriate for the study. However, the sample has fewer female respondents because of less representation in the labour force, a typical characteristic of a developing society.

### 3.2 Analysis

We use statistical methods like Exploratory Factor Analysis (EFA), Confirmatory Factor Analysis (CFA), Structural Equation Modelling (SEM), and Mediation Analysis to study the data.

| S.No. | Demographic Variable | Category | N | Ratio |
|---|---|---|---|---|
| 1. | Gender | Male | 309 | 75.9 |
| | | Female | 98 | 24.1 |
| 2. | Age | 18-24 | 25 | 6.14 |
| | | 25-34 | 135 | 33.2 |
| | | 35-44 | 89 | 21.9 |
| | | 45-54 | 87 | 21.4 |
| | | Above 54 | 71 | 17.4 |
| 3. | Education Level | Less than Bachelor's | 17 | 4.2 |
| | | Bachelor's | 159 | 39.1 |
| | | Master's | 181 | 44.5 |
| | | Doctorate | 47 | 11.5 |
| 4. | Family Monthly Income | Less than 500000 INR | 35 | 8.6 |
| | | 500000-1000000 INR | 108 | 26.5 |
| | | 1000000-2000000 INR | 143 | 35.1 |
| | | More than 2000000 INR | 121 | 29.7 |
| 5. | Family Size | 1-2 | 28 | 6.9 |
| | | 3-4 | 230 | 56.5 |
| | | 5-6 | 114 | 28 |
| | | Above 6 | 35 | 8.6 |
| 6. | Car Ownership | None | 55 | 13.5 |
| | | One | 154 | 37.8 |
| | | Two | 129 | 31.7 |
| | | More than two | 69 | 17 |

Table 1: Demographic Information of Respondents

First, we used exploratory factor analysis to understand the factor structure and determine which measurement items measured the same latent factor. Second, after exploring the factor structure in EFA, we used CFA to test the validity and reliability of the scale and the model fit. The results of the EFA and CFA gave the independent variables to be used in further analysis. Third, we used SEM to study the relationship between the variables and test the hypotheses $H_1$ to H5. Lastly, we used mediation analysis to test hypotheses H6 and study the mediation effect of the components of the theory of planned behaviour.

## 4. Results

This section presents the statistical analysis results. We analysed the data using the methodology described in section 3, and the following results were obtained.

### 4.1 Exploratory Factor Analysis

To understand the structure of the latent constructs in the data, we conducted an exploratory factor analysis (EFA). The dataset includes 33 measurement items that measure consumers' perceptions of the different aspects of electric vehicles. We found the dataset suitable for EFA. The "Kaiser-Meyer-Olkin" (KMO) measure of sampling adequacy was 0.804, indicating that the sample size was adequate. "Bartlett's Test of Sphericity" was significant ($\chi^2$ = 3504.990, df = 210, p < 0.001), giving strong evidence against the correlation matrix being an identity. Hence, the correlation matrix was factorable, and the dataset was suitable for EFA. We used the principal component as the method of factor extraction, eigenvalue (> 1) criteria, scree plot (the elbow was obtained at component 8, from which the curve flattened) and variance explained for factor retention and the varimax method for rotation. We chose the cut-off value for factor loadings to be 0.5, and measurement items with less than 0.5 were not included in the EFA and were removed one by one from our analysis.

As a result, 21 items were condensed into seven factors, explaining seventy per cent of the variance. Since the "Cronbach's alpha" cut-off value is 0.6 in exploratory studies, we dropped one factor for which it was less than 0.6. For the other six factors, it was greater than 0.7, which is acceptable and indicates high internal consistency. Based on the factor loadings, these factors were Optimism (Op), Perceived Behavioural Control (PBC), Attitude (At), Range Anxiety (RA), Innovativeness (In) and Subjective Norms (SN). Thereafter, we used these six factors or latent constructs for further analysis.

### 4.2 Reliability and Validity Analysis

We used the lavaan package in R (Rosseel, 2012) to conduct confirmatory factor analysis (CFA). We used CFA to check the reliability and validity of the scale and constructs. Using "Cronbach's alpha" and composite reliability (CR), we measured the internal consistency of the measurement items for the latent constructs (Fornell & Larcker, 1981). The cut-off value for both indicators is 0.7 (Hair, Sarstedt, Hopkins, & Kuppelwieser, 2014). We found both reliability indicators in the range of 0.76 to 0.86 (Table 4), indicating satisfactory reliability of the measurement items.

We used convergent validity and discriminant validity for validity analysis. Convergent validity measures the degree to which the measurement items are observed to be related, given that they are theoretically related (Chiu & Wang, 2008). We used average variance extracted (AVE) and factor loadings to evaluate convergent validity for each latent construct. We found good convergent validity for the questionnaire. The results are presented in Table 3. All factor loadings, but one, exceeded the cut-off value of 0.6 (Hair, Sarstedt, Hopkins, & Kuppelwieser, 2014). The cut-off value for AVE is 0.5 (Fornell & Larcker, 1981), and the AVE for all latent constructs was between 0.511 and 0.744, indicating good convergent validity.

**Table 3: Factors, Measurements, Factor Loadings and Cronbach's Alpha**

| Factors | Factor Loadings | Cronbach's Alpha | CR | AVE |
|---|---|---|---|---|
| **Optimism** | | | | |
| Future affordability of EV to consumers. | 0.898 | 0.86 | 0.86 | 0.684 |
| Advantages of EV to consumers. | 0.889 | | | |
| Future of charging infrastructure. | 0.675 | | | |
| **Perceived Behavioural Control** | | | | |
| Managing charging and maintenance of EVs. | 0.783 | 0.76 | 0.78 | 0.537 |
| Confident about adopting EVs. | 0.759 | | | |
| Cost of owning an EV is lower than CV. | 0.650 | | | |
| **Attitude** | | | | |
| EV is a unique and innovative product. | 0.570 | 0.79 | 0.80 | 0.512 |
| EV is feasible for daily transportation. | 0.768 | | | |
| EVs suit my lifestyle. | 0.847 | | | |
| EV is an economically viable alternative. | 0.644 | | | |
| **Range Anxiety** | | | | |
| Limited Range impedes EV adoption. | 0.798 | 0.82 | 0.83 | 0.615 |
| EVs are less suitable for long trips. | 0.797 | | | |
| Anxious about running out of battery. | 0.757 | | | |
| **Innovativeness** | | | | |
| Keen to buy a new technology soon. | 0.789 | 0.80 | 0.81 | 0.580 |
| Owning an EV set me apart from others. | 0.775 | | | |
| Willing to invest time to learn about EVs. | 0.718 | | | |
| **Subjective Norms** | | | | |
| Expectations of colleagues affect decisions. | 0.887 | 0.85 | 0.85 | 0.745 |
| Expectations of social circle affect decisions. | 0.838 | | | |

The discriminant validity measures the degree to which the measurements of one construct are more strongly correlated among themselves than with the measures of other constructs (Chiu & Wang, 2008), which it is not intended to be measured. **Table 4** shows the result of the discriminant validity analysis. The diagonal elements of the lower triangular matrix are the square root of the AVEs. For all the latent constructs, all the AVEs are more than the correlation coefficients between any two constructs, indicating good discriminant validity.

**Table 4: Discriminant Validity Analysis**

| Latent Constructs | Op | PBC | At | RA | In | SN |
|---|---|---|---|---|---|---|
| Optimism (Op) | **0.827** | | | | | |
| Perceived Behavioural Control (PBC) | 0.353 | **0.732** | | | | |
| Attitude (At) | 0.314 | 0.537 | **0.715** | | | |
| Range Anxiety (RA) | 0.121 | -0.187 | 0.035 | **0.784** | | |
| Innovativeness | 0.201 | 0.397 | 0.477 | 0.002 | **0.761** | |
| Subjective Norms | 0.211 | 0.528 | 0.452 | -0.104 | 0.454 | **0.863** |

Diagonal elements: Square root of the average variance extracted (AVE).
Off-diagonal elements: Correlation coefficients.

**4.4 Relationship between Latent Constructs**

To study the relationship between latent constructs, we employed Structural Equation Modelling (SEM). We used several parameters and indices to evaluate the model fit. The first parameter was the ratio of Chi-square to the degree of freedom, which should fall between one and three. We found it to be 1.608 for our model. Next, we used the "Tucker-Lewis index" and the "comparative fit index" for goodness of fit. These indices should be greater than 0.9 for a good fit (Schumacker & Lomax, 2015). We also used RMSEA (root mean square error of approximation) and SRMR (standardised root mean square residual). These indices should be less than 0.06 and 0.08, respectively, for a good fit (Schumacker & Lomax, 2015). The SEM model fit was good. We found TLI (0.938) and CFI (0.952) above the threshold and RMSEA (0.036) and SRMR (0.047) within acceptable limits.

Figure 3 presents the relationship between latent constructs obtained from SEM. Optimism (Op) and Innovativeness (In) positively affect the "Theory of Planned Behaviour" components (At, SN, PBC). We also found that Range Anxiety significantly negatively affects Perceived Behavioural Control, consistent with the literature. The Range Anxiety is not significantly related to attitude and subjective norms. We also found significant correlations between latent constructs like Optimism and Innovativeness ($r_{Op, In} = 0.200^{***}$) and between the components of the TPB ($r_{At, SN} = 0.452^{***}$, $r_{SN, PBC} = 0.528^{***}$, $r_{PBC, At} = 0.537^{***}$). There is no theoretically established cause-and-effect relationship between these pairs, and hence, we cannot conclude that one causes the other, but an association exists between these constructs.

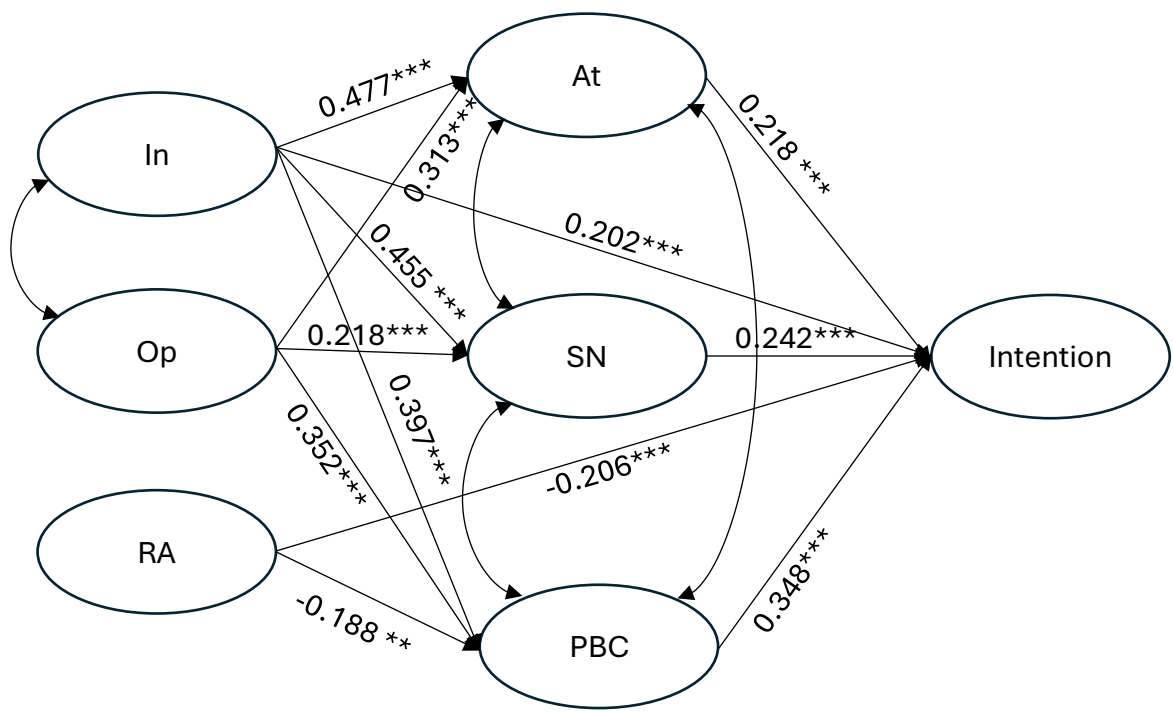

**Figure 3:** The SEM model. 5%, 1%, and 0.1% significance levels are represented by '*', '**', and '***', respectively.

## 4.5 Mediation Analysis

The results of the mediation analysis are shown in Table 5. The mediation analysis partially supports the hypothesis $H_6$. It suggests that the components of the TBP fully mediate the relationship between innovativeness and adoption intention. Full mediation suggests that innovativeness determines adoption intention by shaping the components of the TPB. Innovativeness does not significantly predict adoption intention outside the TPB. The direct effect is completely insignificant when the relationship is mediated by perceived behavioural control. The mediation pathway through perceived behavioural control is the strongest. 63.8% of the total effect is mediated through this pathway, and the proportion is significant, indicating that the impact of innovativeness on adoption intention is predominantly transmitted through PBC. On the other hand, there is no mediation through attitude as the direct effect is significant and the indirect effect is not. There is mediation through subjective norms; however, the proportion mediated is not significant.

| Table 5: Results of Mediation Analysis | | | | |
|---|---|---|---|---|
| Path | Direct Effect(X-Y) | Indirect Effect(X-M-Y) | Total Effect | Proportion Mediated |
| Innovativeness (X) – Attitude (M) – Adoption Intention (Y) | 0.096 (0.047*) | 0.040 (0.100) | 0.136 (<0.001***) | 29.7% (0.280) |
| Innovativeness (X) – Subjective Norms (M) – Adoption Intention (Y) | 0.084 (0.057) | 0.051 (0.024*) | 0.136 (<0.001***) | 37.8% (0.146) |
| Innovativeness (X) – Perceived Behavioural Control (M) – Adoption Intention (Y) | 0.049 (0.216) | 0.87 (<0.001***) | 0.136 (<0.001***) | 63.8% (0.026*) |
| Range Anxiety (X) – Perceived Behavioural Control (M) – Adoption Intention (Y) | -0.112 (0.021*) | -0.044 (0.072) | -0.156 (<0.001***) | 28.4% (0.213) |
| Optimism (X) – Perceived Behavioural Control (M) – Adoption Intention (Y) | -0.054 (0.128) | 0.086 (<0.001***) | 0.032 (0.363) | ---- |

5%, 1%, and 0.1% significance levels are represented by '*','**', and '***', respectively.

The analysis also suggests that the TPB components do not mediate the relationship between range anxiety and adoption intention. The mediation result for mediation through perceived behavioural control is presented in Table 5 (the mediation through other pathways is omitted from the table as they are weak and insignificant). The direct effect is significant even after accounting for perceived behavioural control, and the indirect effect is not; only 28.4% of the total effect is mediated through this pathway, which is not significant. The presence of the direct effect indicates that range anxiety deters adoption intention even if perceived behavioural control improves. This suggests that even when individuals believe that they have control over adopting EVs, they will still be concerned about range anxiety when adopting EVs.

We also found that optimism indirectly affected adoption intention through perceived behavioural control. However, the total effect through this pathway was not significant. The existence of a significant total effect is not necessary to establish mediation (Zhao, Lynch, & Chen, 2010). Hence, we can conclude that optimism affects adoption intention by shaping perceived behavioural control. However, we cannot comment on the proportion mediated as the direct effect is negative and the total effect is not significant. We also found that attitude did not mediate the effect of optimism on adoption intention (the results are not presented in the table). The indirect effect of optimism on adoption intention when attitude is included in the model was not found to be significant (p-value = 0.6008).

## 5. Discussion

This study conceptualises that the consumer perception of electric vehicles is a multi-dimensional concept comprising consumers' perception of cost, maintenance, range, charging infrastructure, government incentives, and other behavioural constructs. This section discusses the key findings and their implication in making consumers' perceptions favourable towards adopting EVs.

First, the statistical analysis results confirm the "Theory of Planned Behaviour" (TPB) postulates. Attitude, Perceived Behavioural Control and Subjective Norms do affect adoption intention. This result follows the literature, where previous studies found that the intention to adopt/purchase is affected by the components of the TPB (Zhang, Bai, & Shang, 2018; Wang, Wang, Li, Wang, & Liang, 2018; Huang & Ge, 2019). Second, we also found that factors (Range Anxiety and Innovativeness) outside the TPB framework impact the electric vehicle adoption intention. The effect of innovativeness on adoption intention is fully mediated by the components of TPB. However, the effect of range anxiety not mediated and directly affects adoption intention. Hence, we have one factor outside the TPB which directly affects adoption intention, and its effect is not explained by the components of the TPB. TPB is a general theory that determines behavioural intentions under all circumstances. However, when it comes to specific objects like the EVs, other factors (other than Perceived Behavioural Control, Attitude and Subjective Norms) might impact behavioural intentions. Literature, including the literature on EV adoption, contains numerous studies (Mohamed, Higgins, Ferguson, & Kanaroglou, 2016; Deka, Dutta, Yazdanpanah, & Komendantova, 2023; Pamidimukkala, Kermanshachi, Rosenberger, & Hladik, 2025) which have extended the theory of planned behaviour (Hu, et al., 2025; Boucif, Nawang, Saadallah, & Mursidi, 2025; Liu, Liu, & Jia, 2025; Czyżewski, Poczta-Wajda, Matuszczak, Smędzik-Ambroży, & Guth, 2025) to include variables that impact behavioural intentions concerning objects in different circumstances.

Third, we found that Optimism about EVs does not impact the intention to adopt them. There exists evidence in the literature to support this result. In the study conducted in the United Kingdom (Salari, 2022), Salari found that consumers' optimism towards EVs did not impact consumers' intention to purchase EVs. Though we see in the previous studies on technology adoption that there is a higher probability of adoption when optimism towards the technology is high (Liljander, Gillberg, Gummerus, & Riel, 2006; Parasuraman, 2000), the role of optimism in EV purchase intention is not well established. A possible explanation for this result could be that even though optimism for the future of EVs is high, other perception variables about EV instrumental attributes like perceived behavioural control and range anxiety make consumers less enthusiastic about EVs in the present when they consider purchasing a vehicle. Hence, to overcome this present bias in consumer perception, there is a need for information dissemination to make them aware of the improvements in EV technology and the incentive structure.

Fourth, we also found that factors like optimism and innovativeness affect the components of the TPB. Optimism directly impacts all the components of TPB, consistent with the literature on technology adoption (Salari, 2022). Optimism does not directly affect adoption intention, but shapes perceived behavioural control. Though there is a significant association of optimism with subjective norms and attitude, the study found no evidence to conclude that these two variables mediate the effect of optimism on adoption intention. We also found that Innovativeness directly affects all three components of the TPB and is significantly correlated with Optimism. Perceived behavioural control and subjective norms mediate the relationship between innovativeness and adoption intention. The mediation primarily takes place through perceived behavioural control, as the mediation pathway through perceived behavioural control is the strongest. 63.8% of the total effect is mediated through this pathway. We also observe a significant correlation between the theory's three components, which, though not reported by many previous studies, is a logical expectation. At this moment, we cannot establish a cause-and-effect relationship between the three components of TPB. Thus, this study adds to the "Extended Theory of Planned Behaviour" literature.

Fifth, since the effect of innovativeness is fully mediated by the components of TPB, any intervention to influence adoption intention should target the mediators. Simply changing innovativeness will not change the behaviour unless one of the mediators is changed. Hence, designing interventions targeting the mediators (perceived behavioural control) is a better strategy. However, interventions can be designed to target range anxiety because components of TPB (perceived behavioural control) only partially mediate the relationship between range anxiety and adoption intention. Therefore, targeting only the mediators will leave concerns unaddressed that may deter adoption intention. Hence, to foster EV adoption, interventions targeting four factors (three from TPB and range anxiety) can be made.

## 6. Conclusion and Future Work

The study adds to the literature on electric vehicle adoption by studying the impact of consumer perception of electric vehicles on their adoption in a tier 2 city in India. It confirms the postulates of the planned behaviour theory and extends the theory by adding new factors in determining behavioural intention towards EV adoption. Though the effect of innovativeness is captured by the components of the TPB, that of range anxiety is not. Hence, there are more factors determining EV adoption than what is indicated by the theory of planned behaviour. We also found that despite high consumer optimism towards the future of EVs, the negative perception impedes consumer decision-making in favour of EVs. Optimism does impact the components of TPB and thereby indirectly impacts adoption intention. Still, the absence of a direct effect reveals the prevalence of negative perception, which needs to be addressed.

The emphasis of the study on consumer perception and the results stresses the fact that technological improvements alone will not lead to the adoption of technology. Given the low level of consumer awareness of the incentives offered by the government for the promotion of EVs, there is a need to make consistent efforts to create a positive perception through intensive information dissemination. Interventions can be made to target mediator variables (components of TPB) and range anxiety. Targeting variables like innovativeness and optimism, whose effects have been explained by the mediating variables, may not be effective. The study can also help design informational interventions to increase the uptake of electric vehicles. Researchers can design informational interventions to target factors like perceived behavioural control, subjective norms, range anxiety, etc. These interventions can aim to modify these factors, like increasing perceived behavioural control or decreasing range anxiety, to improve the intention to adopt EVs. Researchers can also test these interventions in a randomised controlled field experiment and further validate the findings of this study. Future studies can also look into the question of the cause-and-effect relationship between the components of the TPB.

However, this study is not without limitations. We conducted this study in Lucknow (a tier II city in Uttar Pradesh province of India). There is a need to validate this study across geographies and demographics. Also, using self-reported data on adoption intention can be a source of bias. Also, one of the limitations of the TPB is that it focuses on intention to perform a behaviour, which may not translate to actual behaviour, known as the intention-behaviour gap. Respondents can report a favourable intention towards a particular behaviour but may not translate their intention into action. Researchers can use the actual vehicle buying data of the potential buyers to compare the intention and actual decision. Thus, they can validate these findings and bridge the intention-behaviour gap in a long-term study.

Boucif, S. A., Nawang, W. R., Saadallah, O., & Mursidi, A. (2025). Extending the theory of planned behavior in predicting entrepreneurial intention among university students: The role of perceived relational support. *The International Journal of Management Education, 23*, 101168.

Liu, H., Liu, X., & Jia, M. (2025). Revealing the speeding principle based on the extended theory of planned behavior. *Transportation Research Part F: Traffic Psychology and Behaviour, 109*, 1080-1099.

Czyżewski, B., Poczta-Wajda, A., Matuszczak, A., Smędzik-Ambroży, K., & Guth, M. (2025). Exploring intentions to convert into organic farming in small-scale agriculture: Social embeddedness in extended theory of planned behaviour framework. *Agricultural Systems*, 104294.24